\documentclass[preprint,12pt]{elsarticle}

\usepackage{amssymb}

\begin{document}

\title{Very Forward Calorimetry at the LHC - Recent results from ATLAS}

\author{Sebastian N. White\footnote{for the ATLAS Collaboration. }}
\address{Physics Dept., Brookhaven National Lab, Upton, NY 11973, USA}

\begin{abstract}
We present first results from the ATLAS Zero Degree Calorimeters (ZDC) based 
on 7~TeV pp collision data recorded in 2010. 
The ZDC coverage of $\pm \sim350\, \mu$rad about the forward direction makes possible the measurement of neutral
 particles (primarily $\pi$$^{0}$'s and neutrons) over the kinematic region 
$x_F\gtrsim 0.1$ and out to $p_T\lesssim$ 1.2 GeV/c at
 large $x_F$. The ATLAS ZDC is unique in that it provides a complete image of both electromagnetic and hadronic showers.
 This is illustrated with the reconstruction of $\pi$$^{0}$'s with energies of 0.7-3.5 TeV. We also discuss the waveform reconstruction
algorithm which has allowed good time-of-flight resolution on leading neutrons emerging from the collisions despite
the sparse (40 MHz) sampling of the calorimeter signals used.
\end{abstract}

\maketitle


\section{Introduction}

 	Since the ISR, the measurement of leading protons, over a limited kinematic region ($x_F\gtrsim 0.95$) has played an important role in the
 study of strong interactions at colliders. In this region, the phenomenology is characterized by a neutral colorless exchange (``the Pomeron'') which
 is now understood to be more complex in nature than photon exchange in QED. 
 
 	Starting at RHIC, and then
 at HERA, the calorimetric measurement of leading neutrons, which is enabled by colliders with small crossing angle and magnetic separation
 of the beams downstream of the collision, has also become an important tool. This technique is more flexible than that of leading protons, since
 neutrons at essentially all Feynman $x$ values can be detected. The main limitation is the transverse space available for the detector - which typically
 limits the acceptance for forward neutrons (as well as $\pi$$^0$'s and other neutrals) to $p_T\lesssim$ 1 GeV/c. Its utility is, therefore not restricted to 
 diffractive or even soft collisions. It is a more general tool for event characterization, used in the same sense as in Heavy Ion collisions, for
 which it was originally conceived but now for pp and ep collisions as well.
 
 	The analogy to Heavy Ion collisions is an important one, as has been pointed out by Bjorken~\cite{Bj} and Strikman~\cite{Strikman1,Strikman}. At LHC energies
the diffractive slope is increasing, implying a large increase of the proton radius and hence a large difference in the typical impact parameters for soft and hard interactions.
Hence it should be fruitful to identify regimes of central and peripheral collisions where the nature of interactions could differ significantly. Although impact parameter
is treated in event simulation tools, such as PYTHIA8, the tools for analyzing events in terms of this important parameter have been limited\footnote{Moreover fragmentation of the nucleon is not correlated in these models with the impact parameter of the inelastic collision. }. Leading baryons offer one such tool. 

	In pp collisions with the smallest impact parameters at LHC energies, gluon densities are achieved that are as large
as those obtained in central Au+Au collisions at RHIC~\cite{Strikman1,Strikman}. Were tools available for identifying this class of events it would enable further studies of saturation physics, for example.
It is natural to expect that in central collisions the leading baryon will carry a smallish $x_F$ since, in such collisions, two or three valence quarks of the nucleon would be active in the interaction.

\section{Measurements with the Zero Degree Calorimeter}

	The ATLAS ZDC~\cite{Elba}, like the RHIC ZDCs~\cite{RHICZDC}, consists of independently read out modules which sample showers along the beam direction with identical sampling
density and depth (1.2 nuclear interaction lengths/module). Because electromagnetic showers are fully contained in the first module, this module is referred to as EM and photon/hadron discrimination
is simply accomplished with a cut on the energy fraction in the first module.

	The main consideration which drove the design of the ZDC was radiation tolerance since the absorbed dose in the ZDC corresponds to 200 W of continuous energy deposition at full
luminosity and an annual integrated dose of several GigaRad. For this reason, the sampling medium consists of unclad quartz rods and the energy is sampled by Cerenkov light produced in the rods,
predominantly from tertiary relativistic electrons within a well defined angle with respect to the quartz rods. One consequence of the Cerenkov sampling technique is that the hadronic response profile is narrow compared to the actual extent of hadronic showers.

	The main challenge in this radiation hard design has been to achieve good spatial resolution on the transverse position of electromagnetic and hadronic showers ($\sim 0.2, 1.4$ mm, respectively~\cite{Elba}). This is accomplished through the use of a hybrid Shashlik/strip readout technique having different readout granularity in the 1st and 2nd modules.
	
\subsection{Heavy Ion measurements}

	The calibration of collision centralities (i.e. impact parameter, $b$) is best accomplished through the known frequency distribution of collisions (i.e. $\mathcal{L}(b) \propto b\times db$). Any set of observables which change monotonically with respect to $b$ can be mapped to impact parameter, with some dispersion, using this relation. In the PHENIX experiment at RHIC a combination of forward neutron (``spectator'') multiplicity, measured with the ZDC, and charged particle multiplicity in the central detector is used. In order to accomplish this calibration it is critical to obtain a data sample with the minimum bias with 
respect to $b$. This is best done using a ZDC trigger~\cite{MCDPRL,BaltzWhite}.
	
	It is critical for many Heavy Ion analyses to have a minimally biased sample with a well understood calibration over the full range of $b$. Once this is done, for example, modifications of jet distributions in central collisions can be compared directly to the corresponding peripheral, ``pp-like'', collisions with the same center of mass energy ($R_{CP}$) without resorting to special pp collider runs with lower cms energy.

	The reaction plane (i.e. the orientation of $b$) can be measured using directed flow, $v_1$, from the pattern of energy in the ZDC, as was shown by PHENIX~\cite{Hiroshi}. This is all you need to know about experimental Heavy Ion physics. The rest is just physics.
	
\subsubsection{Proton beams}

	The commonly used pp modelling tools (i.e. PYTHIA8 and PHOJET) do not reproduce the features of forward neutron production at RHIC or LHC very well. An alternative approach~\cite{ZDCaccept} has been to synthesize what is known about inclusive distributions from lower energy pp and ep data. The picture that emerges is that, while a forward baryon is always present in each hemisphere, roughly $40\%$ of the time it is a neutron. To describe the doubly-differential distributions, the simplest assumption is that distributions in the 2 hemispheres are, to good approximation, uncorrelated. This model is easily tested against ZDC-based luminosity data at RHIC and LHC. Forward neutron modeling will be further discussed in a separate paper~\cite{Eprint}.
	
\subsubsection{ATLAS ZDC configuration prior to July 2010}

	As discussed in Ref.~\cite{Elba}, the ATLAS ZDC shares space in the LHC neutral absorber (TAN), with LHC luminosity hardware (BRAN), and, for a limited period, the LHCf detector\footnote{ see eg. 
{\tt ab-dep-bi-pm.web.cern.ch/ab-dep-bi-pm/pmwiki/uploads/Activities/
LHCf$\_$ impact.doc}

 }. Prior to the end of July 2010, when the present data were recorded, the front ZDC (nominal EM) module could not be installed since the space was assigned to LHCf. Therefore, for the present data, the ZDC configuration differed from its nominal design in the following way:
\begin{enumerate}
\item
the first (EM) module was preceeded by $\sim$2 radiation lengths of material (BRAN) and, over a small aperture, by the LHCf detector;
\item
the first module had coarser transverse spatial segmentation, since this module was designed for measuring hadronic showers and groups of three quartz rods are read out by a single PMT;
\item
this configuration was shorter by 1.2 nuclear interaction lengths and therefore has somewhat poorer containment of hadronic showers.
\end{enumerate}

	Also, during this period, full offline reconstruction of ZDC data was not yet implemented within the ATLAS offline computing framework, so for the results reported below, a private ``spy buffer'' analysis framework, collecting data directly from the data stream in the ZDC VME crate, was used for energy calibration and $\pi$$^0$ reconstruction. This accounts for the limited statistics in the results presented. ZDC waveform reconstruction and time calibration, using PHOS4 delay scans~\cite{PHOS4}, was developed in {\it Mathematica 7.0} and later ported to C++.

\subsubsection{ZDC waveform reconstruction}

	As discussed in Ref.~\cite{Elba}, although the ZDC trigger signals are brought to the ATLAS control room on short, fast cables, the signals used for digitization, due to cost and infrastructure considerations, use 320 m long ethernet cables  which give poorer timing properties. As a result the signal has a frequency spectrum which rolls over at about 30 MHz ($f_{MAX}$).
		
	In his classic paper deriving the sampling theorem~\cite{Shannon}, Claude Shannon uses the interpolation formula:
\begin{equation}
f(t)=\sum_{n=-\infty}^{\infty}x_n\frac{\sin(\pi(t/T-n))}{\pi (t/T-n)} \;,
\end{equation}
where $x_n$ are sampled values of the waveform at time, $t/T=n$, and shows that once the sampling interval, $T$, is smaller than 1/(2$f_{MAX}$) this formula gives perfect reconstruction of the waveform. In the case of the ZDC there are a total of 7 sampling points spaced 25 nsec apart so we are not in the limit of perfect sampling. Nevertheless, based on our experience~\cite{ATF} with picosecond timing of fast signals, using Tektronix digital scopes, which implement Eq.~(1) on-chip, we find that this formula gives the best possible timing resolution for sparse sampling. Therefore, we decided to use this elegant interpolation formula to reconstruct the time and energy of ZDC waveforms~\cite{ZDCrec}. In this paper we also used 1 nsec delay scan data to map the non-linearities in  response algorithm, when using our digitization in the ATLAS L1calo pre-processor electronics. 

	The timing performance of the ZDC is illustrated in Fig.~1, where we have merged ZDC data $n$-tuples with the corresponding ones from the ATLAS inner tracker. In this particular LHC run, there is evidence for ``satellites'' in the beam bunch structure. This is clear from the inner tracker distribution which shows collisions at $\pm75$ cm as well as at 0. The projection of this plot onto the horizontal axis, which is the time-of arrival of signals measured in the 1st hadronic module on one side, clearly confirms this structure. In this plot, the non-linearity corrections to ZDC timing, derived in Ref.~\cite{ZDCrec}, are intentionally turned off. This plot shows that those corrections could be obtained independently from the satellites, confirming at least a few points of the PHOS4 delay scan. What may be less clear is that Fig.~1 contains a great deal more information about the satellite bunches than could be obtained from the inner tracker distribution by itself. The expected distributions in this plot are derived and animated in a Mathematica notebook~\cite{Mathsim}.
	
\begin{figure}
  \includegraphics[height=.3\textheight]{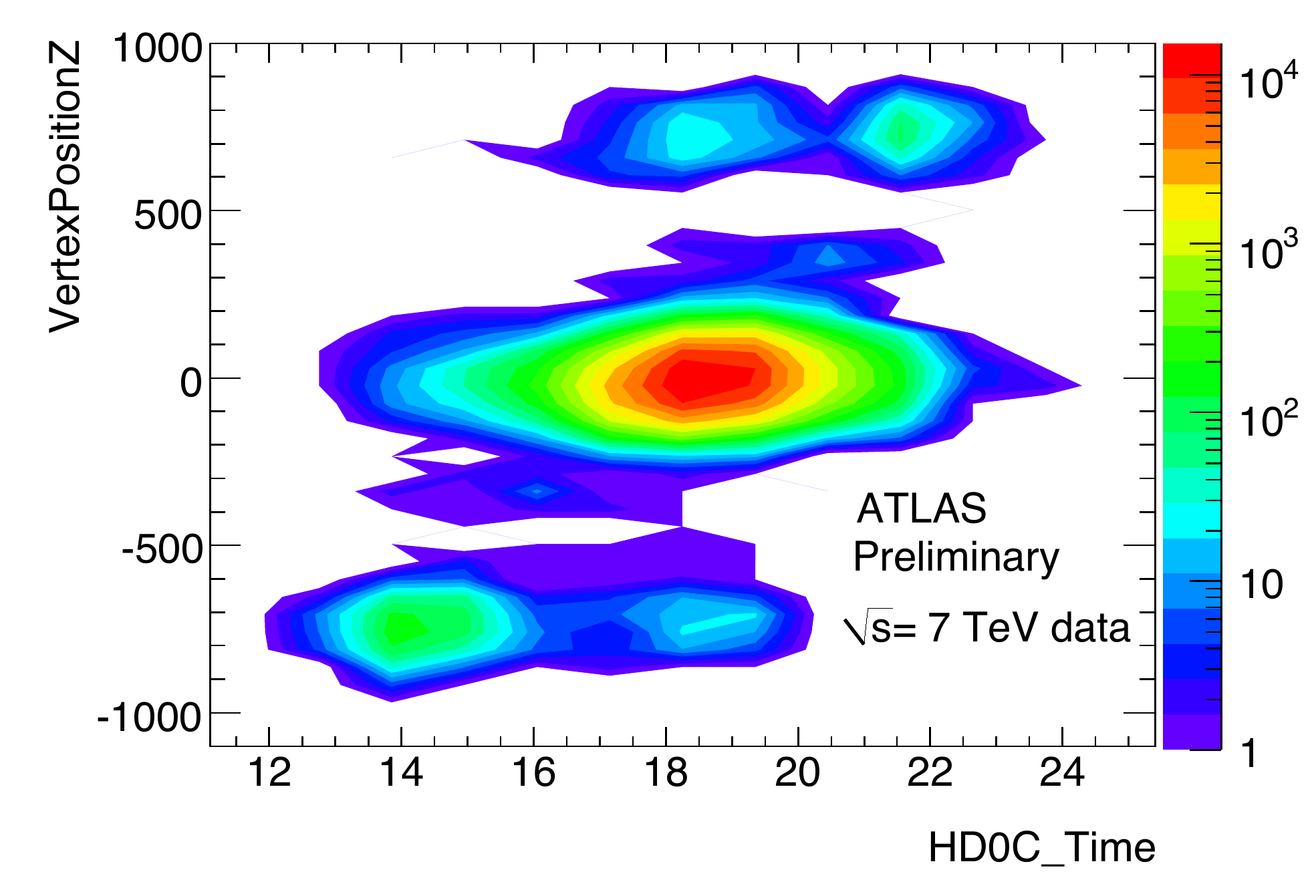}
  \caption{Reconstructed vertex position along the beam direction (in mm) versus the time of arrival (in nanoseconds) of signals at the ZDC 1st hadronic module on the ``C'' side of ATLAS relative to the clock.  The measured time spread ($\sigma_t$) of the satellites at 14.2 and 21.8 nsec is 0.28 and 0.18 nsec, respectively, which is close to the expected spread in arrival time due to the finite bunch length\cite{Mathsim}.The non-linearity corrections measured in Ref.~\cite{ZDCrec} were turned off to make this plot, which confirms these corrections. The main features of the plot are accounted for by the presence of satellite bunches at an interval corresponding to the SPS storage RF frequency.}
\end{figure}

\subsubsection{$\pi$$^0$ Reconstruction}
	
	Energy calibration of the ZDC is far more difficult using pp data than it is with Heavy Ion data, where well resolved single and multiple neutron peaks~\cite{MCDPRL} are easily used for setting
the energy scale. Nevertheless an energy calibration was performed for hadronic and electromagnetic showers, during the period prior to July 2010. Since energy of electromagnetic showers, in particular, is concurrently measured in both the coordinate readout channels and the strip channels (where calibration is simplest) this allows calibration of the coordinate channels as well using Electromagnetic shower candidates. Specific cuts for this analysis were:
\begin{itemize}
\item
all energy is deposited in the first ZDC module
\item
treating the 24 pixel channels as a  4X6 matrix, select events with 2 separated maxima in the amplitude distributions
\item
make a fit for 2 photons (whose showers may be partially overlapped)
\item
check that there are no extra photons
\item
check that total energy of the 2 photons is consistent with the strip energy
\end{itemize}

\begin{figure}
  \includegraphics[height=.3\textheight]{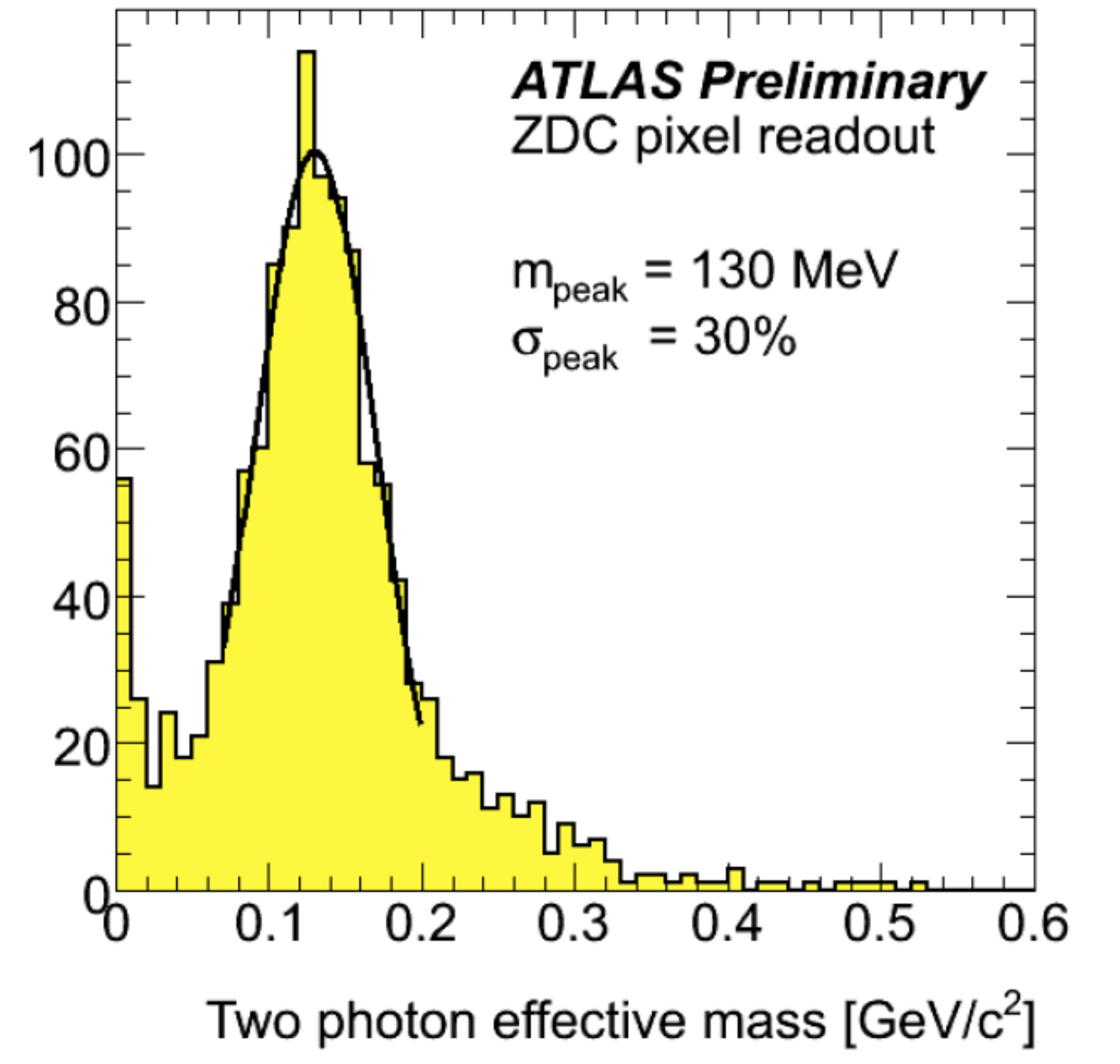}
  \caption{Reconstructed mass distribution for 2 photon candidates, selected as described in the text. The data were obtained from a mixed sample of ATLAS triggers during ATLAS 7 TeV pp running.}
\end{figure}

\section{Outlook}

	At the end of July LHCf was removed and the front modules of the ZDC were installed. Data were then recorded in this configuration for the remainder of the year. It is hoped that we will be able to commission the high resolution EM module coordinate readout also during the January 2011 access.

\section{Acknowledgments}
 	I thank Alessandro Papa and Roberto Fiore for their kind support of my participation in Diffraction 2010. They have dedicated the workshop to the memory of Alexei Kaidalov, who worked on ZDC physics and was supportive of this project. On March 2, 2009- an official BNL "snow day" like the one on which I write this- Volodja Issakov came in to his office at the lab. Later that day he succumbed to a heart attack and ATLAS and ZDC lost an unusually creative detector scientist. Were he alive today the first thing he would want to discuss is how to get even better ZDC performance. Alexei and Volodja would have enjoyed eachother's company.


\end{document}